\documentclass[aps,prd,preprint,nofootinbib,superscriptaddress]{revtex4-1}

\usepackage{graphicx}
\usepackage{epsfig}
\usepackage[hang]{subfigure}
\usepackage{epstopdf}
\usepackage{color}
\usepackage{amsmath}
\usepackage[normalem]{ulem}
\usepackage{xcolor}

\newcommand{\be}{\begin{equation}}
\newcommand{\ee}{\end{equation}}
\newcommand{\ba}{\begin{eqnarray}}
\newcommand{\ea}{\end{eqnarray}}
\newcommand{\beq}{\begin{equation}}
\newcommand{\eeq}{\end{equation}}
\newcommand{\beqa}{\begin{eqnarray}}
\newcommand{\eeqa}{\end{eqnarray}}

\newcommand{\beal}{\begin{aligned}}
\newcommand{\eeal}{\end{aligned}}

\begin{document}

\title{{Thermodynamics of Charged, Rotating, and Accelerating Black Holes}}
\author{Andr\'{e}s Anabal\'{o}n}
\email{andres.anabalon@uai.cl }
\affiliation{Universidad Adolfo Ib\'{a}\~{n}ez; Dep.\ de Ciencias, Facultad de Artes
Liberales - Vi\~{n}a del Mar, Chile.}

\author{{Finnian Gray}}
\email{fgray@perimeterinstitute.ca}
\affiliation{Perimeter Institute, 31 Caroline St., Waterloo, Ontario, N2L 2Y5, Canada}
\affiliation{Department of Physics and Astronomy, University of Waterloo, Waterloo, Ontario, Canada, N2L 3G1}

\author{Ruth Gregory}
\email{R.A.W.Gregory@durham.ac.uk }
\affiliation{Centre for Particle Theory, Durham University, South Road, Durham, DH1 3LE,
UK}
\affiliation{Perimeter Institute, 31 Caroline St., Waterloo, Ontario, N2L 2Y5, Canada}
\author{David Kubiz\v n\'ak}
\email{dkubiznak@perimeterinstitute.ca}
\affiliation{Perimeter Institute, 31 Caroline St., Waterloo, Ontario, N2L 2Y5, Canada}
\affiliation{Department of Physics and Astronomy, University of Waterloo, Waterloo,
Ontario, Canada, N2L 3G1}
\author{Robert B. Mann}
\email{rbmann@uwaterloo.ca}
\affiliation{Department of Physics and Astronomy, University of Waterloo, Waterloo,
Ontario, Canada, N2L 3G1}
\affiliation{Perimeter Institute, 31 Caroline St., Waterloo, Ontario, N2L 2Y5, Canada}

\date{March 18, 2019}

\begin{abstract}
We show how to obtain a consistent thermodynamic description of
accelerating asymptotically AdS black holes, extending our
previous results by including charge and rotation.
We find that the key ingredient of consistent thermodynamics
is to ensure that the system is not over-constrained by including the
possibility of varying the `string' tensions that are responsible for the
acceleration of the black hole, yielding a first law of   full cohomogeneity.
The first law assumes the standard form, with the
entropy given by one quarter of the horizon area and other quantities
identified by standard methods. In particular we compute the mass  in
two independent ways: through a Euclidean action calculation and by
the method of conformal completion. The ambiguity in the choice of the
normalization of the timelike Killing vector can be fixed by explicit coordinate
transformation (in the case of rotation) to the standard AdS form
or by holographic methods
(in the case of charge).
This resolves a long-standing problem of formulating the thermodynamics
of accelerating black holes, opening the way to detailed studies
of their phase behaviour.
\end{abstract}

\pacs{04.20.Jb, 04.70.-s}
\maketitle

\section{Introduction}

Black holes provide us with an invaluable and unique tool for probing the
relationship between quantum physics and the gravitational force.
From the early discoveries that black hole area and surface gravity respectively
behave as thermodynamic entropy \cite{Bekenstein:1973ur, Bekenstein:1974ax}
and temperature \cite{Hawking:1974sw}, a rich and varied range of
thermodynamic behaviour of these objects has been uncovered
\cite{Kubiznak:2016qmn}, providing new insight into the underlying
microscopic degrees of freedom that may be associated with quantum  gravity.

Amongst the panoply of black hole solutions, accelerating black holes have
been somewhat less well understood.  Described by the so-called {\it C-metric}
\cite{Kinnersley:1970zw, Plebanski:1976gy,Dias:2002mi,Griffiths:2005qp}, they
have a  conical singularity along one, or both, polar axes.  Replacing this with
a stress-energy tensor of a finite width cosmic string \cite{Gregory:1995hd}, or
magnetic flux tube \cite{Dowker:1993bt}, gives a concrete interpretation of
the force that accelerates the black holes. Although they have
been used in a variety of settings to demonstrate that the pair creation rate
of black holes is proportional to their entropy \cite{Gregory:1995hd,
Dowker:1993bt,Hawking:1994ii,
Emparan:1995je,Eardley:1995au,Mann:1995vb,Booth:1998pb},
their   thermodynamics has remained perplexing, particularly when charge
and rotation are included.
Conflicting results have appeared in the literature
\cite{Appels:2016uha,Appels:2017xoe,Gregory:2017ogk,
Astorino:2016xiy,Astorino:2016ybm}
concerning the
relationship between the conserved mass and its associated thermodynamic
quantity, the role of conical deficits in the first law, and the relationship
between the action and the free energy.  Recently these inconsistencies were
resolved for the non-rotating uncharged accelerating black hole in
\cite{Anabalon:2018ydc}, by exploring the holographic properties of the
accelerating black hole in AdS with no acceleration horizon, the
\emph{slowly accelerating} black hole \cite{Podolsky:2002nk}.
{Since there is only one  horizon, the system has a unique temperature and thermodynamics is straightforward
to define.}

Here we consider the full thermodynamics of slowly accelerating black holes,
including charge and rotation, and find all relevant thermodynamic variables
showing that both the extended first law \cite{Appels:2017xoe}
\be\label{flaw}
\delta M=T\delta S+\Phi \delta Q+\Omega \delta J-\lambda_+\delta \mu_+
-\lambda_-\delta \mu_- +V\delta P\,,
\ee
and the Smarr relation \cite{Smarr:1972kt}
\be\label{Smarr}
M=2(TS+\Omega J-PV)+\Phi Q\,,
\ee
hold in all cases. We emphasize that both the string tensions $\mu_\pm$, as well
as the cosmological constant, $\Lambda=-8\pi P$, are allowed to vary,
and we derive the associated thermodynamic lengths, $\lambda_\pm$,
and the thermodynamic volume, $V$,  as their respective conjugate quantities.
The scaling properties of $\mu_\pm$ prevent their appearance in the
Smarr relation \eqref{Smarr}.
As we shall see, the inclusion of the tension terms is absolutely crucial for
the laws of thermodynamics to take their natural form.

\section{Generalized C-metric}

We begin by introducing the generalized AdS C-metric solution as derived
from the Pleba\'nski--Demia\'nski  metric \cite{Plebanski:1976gy} in
Boyer--Lindquist type coordinates \cite{Griffiths:2005qp} to include rotation,
charge, cosmological constant $\Lambda=-3/\ell^2$, and the corresponding
gauge potential
\be
\beal
ds^2 &= \frac{1}{H^2}\bigg\{ -\frac{f(r)}{\Sigma}\Big[
\frac{dt}{\alpha}-a\sin^2\!\theta \frac{ d\varphi}{K} \Big]^2 + \frac{\Sigma}{f(r)}dr^2 \\
&\qquad + \frac{\Sigma r^2}{h(\theta)}d\theta^2
+ \frac{h(\theta) \sin^2\!\theta}{\Sigma r^2} \Big[\frac{adt}{\alpha}-(r^2+a^2)
\frac{d\varphi}{K}\Big]^2\bigg\}\,,
\eeal
\label{eq:metric}
\ee
\be
F=dB\,,\qquad B=-\frac{e}{\Sigma r}\Big[\frac{dt}{\alpha}
-a\sin^2\!\theta \frac{d\varphi}{K}\Big]  +\Phi_t dt\,,
\label{eq:elec}
\ee
where we choose
\be\label{Bt}
\Phi_t=\frac{er_+}{(a^2+r_+^2)\alpha}\,,
\ee
so that the gauge potential, defined by $-\xi\cdot B$, where $\xi$ is the
generator of the horizon, {$\xi=\partial_t+\Omega_H\partial_\varphi$}
(where $\Omega_H$ is defined in \eqref{AdSthermo} below),
vanishes at the horizon. The metric functions are given by
\be
\beal
f(r)&=(1-A^2r^2)\bigg[1-\frac{2m}{r}+\frac{a^2+e^2}{r^2}\bigg]
+\frac{r^2+a^2}{\ell^2}\,,\\
{h(\theta)} &= 1+2mA\cos\theta+\bigg[A^2(a^2+e^2)
-\frac{a^2}{\ell^2}\bigg]\cos^2\!\theta\,,\\
\Sigma &=1+\frac{a^2}{r^2}\cos^2\!\theta\,, \qquad
H=1+Ar\cos\theta \,.
\eeal
\ee
As in past work \cite{Appels:2016uha,Appels:2017xoe,Gregory:2017ogk},
we focus on the slowly accelerating black hole that has no acceleration
horizon, i.e.\ the only relevant zero of $f$ being the black hole horizon.
The conformal factor $H$ then
sets the location of the boundary at $r_{\mathrm{bd}}=-1/A\cos\theta$.

It is worth remarking on the various parameters appearing in the metric:
$m$, $a$, $e$ and $\ell$ are the usual parameters representing the mass,
rotation, charge of the black hole and the cosmological constant respectively.
Loosely speaking, $A$ encodes the acceleration of the black hole, and $K$
the conical deficit of the spacetime. Often, the $K$ factor is absorbed in the
azimuthal coordinate, which in turn would have an arbitrary periodicity,
usually fixed by a regularity condition at one of the poles. We include the
explicit factor of $K$ so that the periodicity of $\varphi$ is fixed at $2\pi$.
The conical deficit at each axis is then found by checking the behaviour
of the $\theta-\varphi$ part of the metric near $\theta=\theta_\pm=0,\pi$
respectively:
\be
ds_{\theta,\varphi}^2 \propto d\theta^2 +  h^2(\theta) \sin^2\theta
\frac{d\varphi^2}{K^2} \sim d\vartheta^2 + (\Xi\pm2mA)^2 \vartheta^2 d\varphi^2\,,
\ee
where $\vartheta_\pm = \pm(\theta-\theta_\pm)$ gives a local
radial coordinate near each axis, and
\be
\Xi = 1 - \frac{a^2}{\ell^2} + A^2 (e^2+a^2)\,.
\ee
The definition of a conical singularity is that the circumference of a
circle, ${\cal C}=\Delta \varphi \sqrt{g_{\varphi\varphi}}$, is
not equal to $2\pi$ times its proper radius, ${\cal R}
=\vartheta\sqrt{g_{\theta\theta}}$. The deficit angle
is this discrepancy: $\delta = 2\pi -{\cal C}/{\cal R}$.
This is the reason
for the parameter $K$: by fixing the periodicity of $\varphi$ as
$\Delta \varphi = 2\pi$, the deficits at each axis are now given explicitly
in terms of the parameters appearing in the metric.
Since a deficit angle is
typically interpreted as due to the presence of a cosmic string, one
deduces that the tensions of the strings along each axis, $\mu_\pm
= \delta_\pm/8\pi$, are:
\be\label{eq:deficits}
\mu_\pm =\frac{1}{4}\bigg[1-\frac{\Xi \pm 2mA}{K}\bigg]\,.
\ee
We therefore see that acceleration is due to a mismatch of conical
deficits from the North to South pole: $\mu_--\mu_+ = mA/K$,
whereas $K$ tracks an overall deficit in the spacetime:
$\bar{\mu} = (\mu_++\mu_-)/2 = \frac14(1-\Xi/K)$.

Thus far we have related all parameters in the metric to physical
charges. We emphasize that  $\alpha$, the parameter
rescaling $t$, is not a new parameter:
since $t$ is a noncompact coordinate it is not a physical parameter.
Rather it is a choice of gauge, anticipating that the $t$ coordinate natural
to the $(t,r)$ section of a black hole metric is not, in fact, the correctly
scaled coordinate when viewed from the perspective of an observer
at infinity (for the non accelerating Kerr-AdS, the presence of this
rescaling was first noted in \cite{Gibbons:2004ai}.)
We insert $\alpha$  at this stage to anticipate
this rescaling -- see \eqref{alp-eq} below.

To summarize, the general solution above is described by six independent
parameters $\{m, a, e, \ell, K, A\}$, which map to six physical charges
$\{M, J, Q, P, \mu_+,\mu_- \}$, that all appear on the RHS of the First Law
\eqref{flaw}. As a consequence the first law has ``full cohomogeneity''.

Finally, in order for the metric \eqref{eq:metric} to be well defined, and correspond
physically to a slowly accelerating black hole in the bulk, we must restrict
the range of parameters as follows:
\begin{enumerate}
\item
For the obvious interpretation of
the $\theta$-coordinate, we must have $h(\theta)>0$ on $[0,\pi]$, which implies
\be
mA<\begin{cases}
\frac{1}{2}\Xi\quad\mbox{for}\quad \Xi\in(0,2]\,,\\
\sqrt{\Xi-1}\quad  \mbox{for}\quad \Xi>2\,.
\end{cases}
\ee
(Other ranges of $\theta$ where $h$ is positive  yield, for example,
hyperbolic black holes \cite{Hubeny:2009kz}.)

\item
Since the range of $r$ is constrained by the conformal factor,
we have\footnote{
Note that when $\cos\theta < 0$, $1/r$ crosses the origin and so the
boundary is situated `beyond infinity'.} $A\cos\theta \leq  1/r  \leq 1/r_+$,
and to  satisfy the requirement of  {slow} acceleration, we must require that
$f(-1/A\cos\theta)$ has no roots.

\item
Finally, we must make sure that the
spacetime admits a black hole in the bulk, that is $f(r)$ has to have at least
one root in the range $r\in(0,1/A)$.
\end{enumerate}

Once we establish that we are within these
allowed parameter ranges, our solution describes a
single accelerating black hole that is suspended by strings at a constant radius
in asymptotically AdS space. We refer the reader to reference
\cite{Podolsky:2003gm} for the (rather non-trivial and $\theta$-dependent)
Penrose diagram of these spacetimes.

\section{Slowly Accelerating Black Hole Thermodynamics}

We begin by stating the thermodynamic charges and potentials for
the accelerating black hole:
\be
\beal
M&= \frac{m(\Xi+a^2/\ell^2)(1-A^2 \ell^2\Xi)}{K\Xi\alpha(1+a^2A^2)}\,,\\
T&= \frac{f'_+ r_+^2}{4\pi\alpha(r_+^2+a^2)}\,, \qquad
S=\frac{\pi(r_+^2+a^2)}{K(1-A^2r_+^2)}\,,\\
Q&= \frac{e}{K}\,,\quad \Phi=\Phi_t=\frac{er_+}{(r_+^2+a^2)\alpha}\,,\\
J& =\frac{ma}{K^2}\,,  \quad \Omega=  \Omega_H-\Omega_\infty\,
=\left ( \frac{Ka}{\alpha(r_+^2+a^2)}\right ) -
\left ( -\frac{aK(1-A^2\ell^2\Xi)}{\ell^2\Xi \alpha(1+a^2A^2)}\right) \,,\\
P &= \frac{3}{8\pi \ell^2} \,, \quad
V = \frac{4\pi}{3K\alpha} \left [ \frac{r_+(r_+^2 + a^2)}{(1-A^2 r_+^2)^{2}}
+ \frac{m[a^2(1-A^2\ell^2 \Xi) + A^2 \ell^4 \Xi (\Xi+a^2/\ell^2)]}{(1+a^2 A^2) \Xi} \right]\,,\\
\lambda_\pm &= \frac{r_+}{\alpha(1\pm Ar_+)} - \frac{m}{\alpha}
\frac{[\Xi + a^2/\ell^2 +   \frac{a^2}{\ell^2} (1-A^2\ell^2 \Xi)]}{(1+a^2 A^2)\Xi^2}
\mp \frac{A \ell^2 (\Xi +  a^2/\ell^2 )}{\alpha(1+a^2A^2)}\,,
\eeal
\label{AdSthermo}
\ee
which, together with the tensions $\mu_\pm$ defined in \eqref{eq:deficits}, satisfy
both the first law \eqref{flaw} and the Smarr relation \eqref{Smarr}, provided
we set
\be\label{alp-eq}
\alpha=\frac{\sqrt{(\Xi+a^2/\ell^2)(1-A^2 \ell^2\Xi)}}{1+a^2A^2}\,.
\ee
Let us now turn towards the derivation of these quantities.

\medskip
\noindent$\bullet$ $T$ and $S$:

The most straightforward of the above parameters is the entropy, given by
one quarter of the horizon area \cite{Herdeiro:2009vd},
\be
S=\frac{\mbox{Area}}{4}=\frac{r^2+a^2}{4K}\int\frac{\sin\theta}{H^2}
d\theta d\varphi\Big|_{r=r_+}=\frac{\pi(r_+^2+a^2)}{K(1-A^2r_+^2)}\,.
\ee
Next, the Hawking temperature of the black hole, $T$, can be
obtained as usual by regularity of the Euclidean section\footnote{
Note that when Wick rotating Kerr $ia$ appears in the metric.  However
to check regularity at the horizon one must approach in a locally
co-rotating frame: $(r_+^2+a^2)\varphi/K = -ia\tau/\alpha$. This
results in a manifestly $\theta-$independent answer following the
usual procedure. Alternately, using the zeroth law, the temperature
is constant on the horizon, and hence can be evaluated at $\theta=0$ for simplicity.}
\be
ds^2_{\tau,r} \propto \frac{r^2\Sigma}{r^2+a^2} \left [
\frac{r^2f }{(r^2+a^2)} \frac{d\tau^2}{\alpha^2}
+\frac{dr^2}{r^2f/(r^2+a^2)} \right] \quad \Rightarrow \quad
T=\frac{f'r_+^2}{4\pi \alpha (r_+^2+a^2)}\,.
\ee

\medskip
\noindent$\bullet$ $Q$ and $\Phi$:

The electric and magnetic charges of the black hole, $Q$ and $Q_m$,
can be found by the standard Gauss' law integral
\be\label{Q}
Q=\frac{1}{4\pi} \int * F=\frac{e}{K} \qquad Q_m=\frac{1}{4\pi} \int  F=0\,.
\ee
The conjugate electrostatic potential, $\Phi$, is a little more interesting.
The usual procedure is to define the bulk potential $B$ so that the
corresponding electrostatic potential $-\xi\cdot B$ vanishes on the horizon of
a black hole, and its value at infinity then gives the thermodynamic potential.
Here, however, we have the problem that the potential and electric field are
not constant at the boundary due the acceleration giving a nontrivial
$\theta-$dependence, so we cannot simply read off the potential there.
Instead, as already noted in \eqref{eq:elec} and \eqref{Bt}, 
we set
the electrostatic potential $-\xi\cdot B$ to be zero on the horizon,
and define the thermodynamic potential using
the Hawking--Ross prescription \cite{Hawking:1995ap}
\be
\Phi=\frac{1}{4\pi Q\beta}\int_{\partial M} \sqrt{h}n_a F^{ab} B_b\,,
\label{Icharge}
\ee
where $\beta=1/T$ is the inverse periodicity in Euclidean time, and
$n^a$  is the outward pointing unit normal to $\partial M \equiv \{H=0\}$.
This computation simply gives $\Phi=\Phi_t$, as in \eqref{Bt}.

\medskip
\noindent$\bullet$  $J$ and $\Omega$:

The angular momentum $J$  can be computed (for
example) by using the conformal method \cite{Ashtekar:1999jx,Das:2000cu}.
The idea is to perform a conformal transformation on the metric,
$\bar{g}_{\mu\nu} = {\bar{H}^{2}g_{\mu\nu}}$,
to remove the divergence near the boundary, then to integrate a conserved
current associated with the Killing vector $\xi$, to get the corresponding
conserved charge:
\be
\textsf{Q}(\xi)=\frac{\ell}{8\pi}\lim_{\bar{H} \rightarrow 0}\oint
\frac{\ell^{2}}{\bar{H}}N^{\alpha }N^{\beta }
\bar{C}^{\nu}{}_{\alpha \mu \beta }
\xi _{\nu }d\bar{S}^{\mu }\,,
\ee
where $\bar{C}^{\mu }{}_{\alpha \nu \beta }$ is the Weyl tensor of the
conformal metric and
\be
d\bar{S}_{\mu }=\delta^t_\mu
\frac{\ell^2(1+a^2A^2\cos^4\!\theta)}{\alpha K} d(\cos\theta) d\varphi
\ee
is the spacelike surface element
tangent to $ \bar{H}=0$, with the normal
to the boundary  $N_{\mu }=\partial _{\mu }{\bar{H}}$.
The charge $\textsf{Q}(\xi)$  is independent of the choice of conformal
completion even though the conformal completion is not unique.
The choice $\bar{H}=\ell H r^{-1} $ provides a convenient
conformal completion, smooth as $A\to 0$.

The angular momentum \eqref{AdSthermo} is given by
$\textsf{Q}(-\partial_\varphi)$. As is well known \cite{Gibbons:2004ai}, the angular
potential of a rotating black hole in AdS is not simply the angular momentum
at the horizon in Boyer--Lindquist coordinates, but must be corrected to allow for
the relative rotation with the boundary. In general, the angular velocity of the
``Zero-Angular-Momentum-Observer'' any radius $r$ is
\be\label{omegagen}
\Omega_r=-\frac{g_{t\varphi}}{g_{\varphi\varphi}}
=\frac{Ka[h(r^2+a^2)-fr^2]}{\alpha[h(r^2+a^2)^2-fa^2r^2\sin^2\!\theta]}\, .
\ee
Setting $r=r_+$, this yields $\Omega_H$ in \eqref{AdSthermo} on the horizon
but is not constant on the boundary $H=0$. However, one can define
\be
\Omega_\infty=-\frac{aK(1-A^2\ell^2\Xi)}{\ell^2\Xi \alpha(1+a^2A^2)}\,
\ee
by evaluating \eqref{omegagen}  at conformal infinity with $m = 0$
(and $\cos\theta=1$ for the charged black hole).
Similar to the Kerr-AdS case, it is the difference between the two,
$\Omega=\Omega_H-\Omega_\infty$ that enters the first law of
black hole thermodynamics.
\newpage

\noindent$\bullet$ $M$:

Moving to the thermodynamic mass, we used three methods to find
the expression in \eqref{AdSthermo}. First, the conformal method with
\be
M= \textsf{Q}(\partial_t+\Omega_\infty \partial_\varphi)\,,
\ee
yields the expression in \eqref{AdSthermo}.

We can check this result by computing the  Gibbs free energy
${\cal G}=I/\beta={\cal G}(T, \Omega, \Phi)$ from the action
\begin{align}
&I=\frac{1}{16\pi}\int_{M}d^{4}x\sqrt{g}\left[ R+\frac{6}{\ell^{2}}-F^2\right]
+ \frac{1}{8\pi}\int_{\partial M}d^{3}x\sqrt{h}\left[\mathcal{K}
-   \frac{2}{\ell} - \frac{\ell}{2}\mathcal{R}\left( h\right) \right]\,,
\label{action}
\end{align}
where  $\mathcal{K}$ and $\mathcal{R}\left( h\right)$ are respectively
the extrinsic curvature and Ricci scalar of the boundary.
Computing this free energy is a bit lengthy, but straightforward, and yields
\be
 {\cal G}=\frac{m(1-a^2A^2-2A^2\ell^2\Xi)}{2\alpha K (1+a^2A^2)}
-\frac{r_+(a^2+r_+^2)}{{2}\alpha K \ell^2(1-A^2r_+^2)^2}
{-}\frac{e^2r_+}{2K\alpha(a^2+r_+^2)}\,,
\ee
which satisfies ${\cal G}=M-\Omega J -TS- {\Phi Q}$ \cite{Caldarelli:1999xj}, with
the thermodynamic variables in \eqref{AdSthermo} as required.

The third cross-check is to use integration of the holographic stress tensor.
In the fixed potential ensemble (without the $I_Q$ term), varying
the action \eqref{action} (for generic variations that can include
boundary diffeomorphisms \cite{Papadimitriou:2005ii}) gives
\be
\delta I=-\frac12\int_{\partial M}\mathcal{T}_{ab}
\delta h^{ab}\sqrt{-h}~d^{3}x,
\ee
where
\begin{equation}
8\pi {\cal{T}}_{ab} = \ell \mathcal{G}_{ab}\left(h\right)
-\frac{2}{\ell}h_{ab}-\mathcal{K}_{ab}+h_{ab}\mathcal{K}\,.
\label{Tabdefn}
\end{equation}
In order to compute the quantities appearing in this
expression, we require a systematic expansion of the metric near
the boundary.
This is typically done via a set of asymptotic
Fefferman-Graham coordinates, in which the geometry takes a
standard format
\be
ds^2= \frac{\ell^2}{z^2} dz^2 +
z^{-2} \left ( \gamma_{ab}^{(0)} + z^2 \gamma^{(2)}_{ab}
+ ...\right )  dx^a dx^b\,.
\label{FGmetric}
\ee
The transformation between the C-metric and FG coordinates is
then written as an expansion in $z$, and expressions computed order by
order. This process is straightforward but lengthy,
and determines the $\gamma^{(n)}_{ab}$ in \eqref{FGmetric} above.
Since $h_{ab} = \gamma_{ab}^{(0)} /z^2 + \gamma_{ab}^{(2)} +\dots$,
it is now straightforward to compute ${\cal K}_{ab}$
in \eqref{Tabdefn} leading to
\begin{equation}
8\pi {\cal{T}}_{ab} = \ell \mathcal{G}_{ab}
+\frac{1}{\ell}\left [ \gamma^{(2)}_{ab} + \frac{3z}{2} \gamma^{(3)}_{ab}
- \gamma^{(0)}_{ab} {\gamma^{(0)}}^{cd}
\left ( \gamma^{(2)}_{cd}  + \frac{3z}{2} \gamma^{(3)}_{cd}
\right)\right]\,.
\end{equation}
Inserting the computed values of $\gamma^{(n)}$, allows the computation of
the expectation value of the energy momentum of the CFT$_{3}$,
\begin{equation}
\left\langle \mathcal{T}_{ab}\right\rangle =\lim_{z\to0}
\frac{\mathcal{T}_{ab}}{\ell z}\,,
\end{equation}
yielding a relativistic fluid whose energy density can be integrated to
give the mass. The details on this procedure are given
in Appendix \ref{app:hol}.

\medskip
\noindent$\bullet$ $\alpha$:

The crucial step in formulating the correct thermodynamics is to
determine the normalization $\alpha$ in \eqref{alp-eq}. For the uncharged
and non-rotating black hole, this is straightforward to determine, as
described in
\cite{Anabalon:2018ydc}: in the limit of vanishing black hole mass and string
tensions, the spacetime becomes pure AdS in Rindler coordinates.
The coordinate transformation between Rindler and global AdS then
determines the rescaling of the time coordinate of the C-metric.

Alternatively, note that in this same limit, the boundary metric becomes
(see  eq.\ \eqref{gamma0} in Appendix \ref{app:hol})
\be
ds_{(0)}^2 = \omega^2 \left [ - \frac{d\tau^2}{\ell^2}
+ \frac{\alpha^2 dx^2}{(1-x^2)(1-A^2\ell^2(1-x^2))}
+ \frac{\alpha^2 (1-x^2) d\varphi^2}{(1-A^2\ell^2(1-x^2))}\right]\,.
\ee
The latter two terms should correspond to a unit $S^2$, so we set
\be
\sin^2\vartheta =  \frac{\alpha^2 (1-x^2)}{(1-A^2\ell^2(1-x^2))}\,.
\ee
Computing the $\vartheta$ component of the metric then gives
\be
\frac{\alpha^2 \cos^2\vartheta d\vartheta^2}
{\alpha^2-(1-A^2\ell^2) + (1-A^2\ell^2)\cos^2\vartheta}
\ee
hence $\alpha^2 = 1-A^2\ell^2$.

For non-zero rotation similar approaches apply. Set $m=e=0$
in \eqref{eq:metric}, and $K=1+a^2A^2-a^2/\ell^2 $ so that there
are no conical deficits present. We then rewrite the metric in a
non-rotating frame by setting $\varphi=\tilde \phi+\Omega_\infty t$,
thus eliminating the $g_{t {\tilde{\phi}}}$ component of the metric:
\be
ds^2=\frac{1}{H^2} \Bigl( -\frac{\tilde f \tilde h dt^2}{\alpha^2 \tilde \Xi}
+\frac{r^2\Sigma dr^2}{(r^2+a^2)\tilde f}+\frac{r^2\Sigma d\theta^2}{\tilde h}
+\frac{\sin^2\!\theta(r^2+a^2)d\tilde \phi^2}{\tilde \Xi}\Bigr)\,,
\ee
where
\be
\tilde f =1-A^2r^2+\frac{r^2}{\ell^2}\,,\quad
\tilde h=1+\left(a^2A^2-\frac{a^2}{\ell^2} \right )\cos^2\!\theta\,,\quad
\tilde \Xi=K=1+a^2A^2-\frac{a^2}{\ell^2}\,.
\ee
This is to be compared with the AdS metric:
\be
ds^2=-\Bigl(1+\frac{R^2}{\ell^2}\Bigr)dt^2+\frac{dR^2}{1+\frac{R^2}{\ell^2}}
+R^2(d\Theta^2+\sin^2\!\Theta d\tilde \phi^2)\,.
\ee
Equality of the $g_{tt}$ and $g_{\tilde\phi \tilde\phi}$ components in each
case determines $R(r,\theta)$ and $\Theta(r,\theta)$:
\be
R\sin\Theta=\frac{\sin\theta \sqrt{r^2+a^2}}{H \sqrt{\tilde \Xi}}\,,\quad R=\ell \sqrt{\frac{\tilde f \tilde h}{H^2 \alpha^2 \tilde \Xi}-1}\,.
\ee
Then imposing that
the remaining metric components agree
yields $\alpha=\sqrt{1-A^2\ell^2}$, as in the non-rotating case.

\medskip
\noindent$\bullet$ $V,\lambda_\pm$:

The remaining thermodynamic variables are the pressure,
string tensions, and their dual potentials. The definition of the charges
$P, \mu_\pm$ are fixed in terms
of the physical quantities: the cosmological constant determines
$P = 3/8\pi \ell^2$, and the tensions are fixed by the conical
deficits \eqref{eq:deficits}. The conjugate potentials, the thermodynamic
volume and lengths are then determined from the first law \eqref{flaw}.
This is computed by first looking at how the location of the black hole
event horizon changes as we vary the metric parameters
\be
\frac{\partial f}{\partial r_+} \delta r_+ +
\frac{\partial f}{\partial m} \delta m +
\frac{\partial f}{\partial e} \delta e +
\frac{\partial f}{\partial a} \delta a +
\frac{\partial f}{\partial \ell} \delta \ell +
\frac{\partial f}{\partial A} \delta A +
\frac{\partial f}{\partial K} \delta K =0\,.
\ee
Next, $\delta r_+$ (conveniently multiplied by $f'_+\propto T$)
can be rewritten in terms of $\delta S$,
$\delta a$,  $\delta A$,  and $\delta K$. This now gives
$T\delta S$ in terms of $\delta m$,  $\delta e$,  $\delta a$
$\delta \ell$,  $\delta A$,  and $\delta K$. We then use the
expressions for $M$, $Q$, $J$, $P$ in \eqref{AdSthermo}
and $\mu_\pm$ in \eqref{eq:deficits} to transform this into
the form of the first law. This procedure results in the expressions
for $V$ and $\lambda_\pm$ in \eqref{AdSthermo}.

\section{On thermodynamics of the asymptotically flat C-metric}

It is also interesting to consider what happens if there is an acceleration horizon.
For simplicity, we concentrate on the asymptotically flat limit of the accelerated
black holes\footnote{Alternatively, one could consider the `fast accelerating'
AdS black holes. Such spacetimes have both accelerated and cosmological
horizons extending all the way to conformal infinity and the standard holographic
and conformal methods are not so readily applicable for their study.}, obtained by
setting $\Lambda=0$ in \eqref{eq:metric}.  It describes  a pair of accelerating
black holes separated by an
acceleration horizon, and we note  that its conformal structure
is  analogous to de Sitter spacetime, but with future/past infinity being
null at the poles $\theta=0,\pi$ and otherwise spacelike \cite{Griffiths:2006tk}.

Given this clear physical distinction, finding the
thermodynamic variables is not a matter of taking the limit $\ell \to \infty$
for each quantity in \eqref{AdSthermo};  indeed   such a limit does not
exist. Rather we concentrate on an observer in the static patch of
\eqref{eq:metric}, between one black hole and its  acceleration horizon.
Despite the fact that charged rotating black holes emit gravitational and
electromagnetic radiation at infinity \cite{Podolsky:2003gm,Podolsky:2009ag},
observers using the metric \eqref{eq:metric}  should be able to  undertake
standard thermodynamic investigations, since the radiation escapes into a
region inaccessible to this observer \cite{Boulware:1980}.
Of course, due to the presence of two horizons such a system is in
general out of equilibrium.
We expect that similar to the de Sitter case, one may study the
thermodynamics of each horizon separately \cite{Dolan:2013ft}.

Leaving the thermodynamics of the accelerated horizon for future study
(see \cite{Bianchi:2013rya,DeLorenzo:2017tgx}), with this in mind we find
\be
\beal
M &=  \frac{m(1-a^2A^2)}{K\alpha(1+a^2A^2)}\,, \qquad
T=\frac{f'_+ r_+^2}{4\pi\alpha(r_+^2+a^2)}\,, \qquad
S =\frac{\pi(r_+^2+a^2)}{K(1-A^2r_+^2)}\,,  \\
\alpha &= \frac{\sqrt{(1-a^2A^2)\Xi}}{1+a^2A^2}\,,\qquad
Q=\frac{e}{K}\,, \qquad \Phi=\frac{er_+}{(a^2+r_+^2)\alpha}\,,
\\
J&=\frac{ma}{K^2}\,, \qquad
\Omega=\Omega_H-\Omega_\infty =
\frac{aK}{\alpha(r_+^2+a^2)}\,-
\frac{aA^2K}{\alpha(1+a^2A^2)}\,,\\
\lambda_\pm &= \frac{r_+}{\alpha(1\pm Ar_+)} - \frac{MK}{\Xi}
\mp \frac{a^2 A}{\alpha(1+a^2 A^2)}\,,&
\eeal
\label{flatthermo}
\ee
form a consistent set of thermodynamic parameters for the $\Lambda=0$
charged accelerating black hole, satisfying (with $P=0$) both the first law
\eqref{flaw} and the Smarr relation \eqref{Smarr} at the black hole horizon,
where $\Xi=(1+a^2A^2+e^2A^2)$.
The quantity $M$ was computed via a Komar-like integral over $\theta$
and $\varphi$ at conformal infinity, using a Killing vector
$\partial_t$ in a rotation subtracted frame $\varphi\to \varphi+\Omega_\infty t$.
Note that it is not possible to compute  $V$ in this case, as no smooth limit
exists for $V$ as $P\to 0$, similar to what happens for black holes in $d=7$
gauged supergravity \cite{Cvetic:2010jb}.

A much greater measure of caution should be  taken before the formulae
in \eqref{flatthermo} are accepted as thermodynamic quantities.  In computing
the mass the Killing vector $\partial_t$ was used; this is a boost rotation Killing
vector rather than the static vector at infinity. Moreover, spatial infinity is not
well described in the static coordinates---it only corresponds to $\theta=\pi/2$
in these coordinates, calling into question the validity of the Komar integration.
Third, the above quantities satisfy the first law only provided this Killing vector
is normalized by $\alpha$ above. Whereas there is a good reason to do this in
the asymptotically AdS case studied in the main text, such a normalization
remains a question in the asymptotically flat case.  Finally, the physical status
of the thermodynamic relations is not clear given
the presence of radiation \cite{Podolsky:2003gm,Podolsky:2009ag};
indeed even in the slowly accelerating AdS case
radiation patterns have been computed \cite{Podolsky:2003gm}.

\section{Discussion and summary}

To summarize, we have formulated the consistent thermodynamics of
slowly accelerating AdS black holes with charge and rotation.
The crucial aspects of our result are the computation of the thermodynamic
mass and the normalization of the timelike Killing vector.  We computed
the mass directly via the conformal method, and backed up this result by
checking it against the free energy from a calculation of the action and
(for zero rotation)
by calculating the holographic energy momentum tensor and its associated energy.
The mutual agreement of these methods is not obvious,
as the Dirichlet boundary conditions no longer apply to the case in question
\cite{Hollands:2005wt}.
The normalization of the timelike Killing vector was cross-checked by finding explicit coordinate transformations and (for zero rotation)
by requiring the variation of the action (including
boundary counterterms) to vanish.

The key insight in our approach is to consider the extended first law
\eqref{flaw} with the additional terms that allow us to vary the tension.
This is in direct contrast to the approach presented in \cite{Astorino:2016xiy}
(see also \cite{Astorino:2016hls}) where one tension is fixed (at zero) and all
remaining parameters are permitted to vary. This is perplexing, as a first law
without varying tension then has reduced cohomogeneity, i.e.\  fewer
thermodynamic parameters than the number of parameters being varied in
the geometry.  Given this mismatch of parameter space, we suspect a
lack of uniqueness, or a hidden constraint.

We have also  proposed a set of thermodynamic parameters for the
$\Lambda=0$ charged and accelerating black hole spacetime. This
spacetime has an acceleration horizon, but (as is the case with other
black holes) the first law and Smarr formula hold at the black hole horizon.

Our results pave the way for a proper analysis of the phase behaviour
of accelerating black holes and raise a number of interesting questions
for future study. Standard thermodynamic analysis  seems to be possible
for observers using the metric \eqref{eq:metric} (e.g.\ \cite{Zhang:2018hms})
despite the fact that charged rotating black holes emit gravitational and
electromagnetic radiation at infinity \cite{Podolsky:2003gm,Podolsky:2009ag},
since the coordinates from which we obtained our thermodynamic parameters
are analogous to those of an observer comoving with an accelerating charged
object, who sees no radiation because it escapes into a region inaccessible
to this observer \cite{Boulware:1980}.
It is an interesting future project to obtain a deeper understanding of this
relationship. Similar remarks apply to the $\Lambda \geq 0$ cases as well
as the fast accelerating AdS case, though the loss of thermodynamic
equilibrium (duel to the presence of additional horizons) introduces
new complications. The quantities in \eqref{flatthermo} were with regard
to one accelerated black hole in the static patch. It remains an interesting
question whether one should not rather treat the system of two black holes
as a whole, reminiscent of the mass dipole, and consider the associated
boost mass \cite{Dutta:2005iy}.

It is also of interest to obtain a better understanding of these black
holes from a holographic perspective. While we have checked our
expression for the mass using this approach, a full understanding of the
dual fluid interpretation for the charged and rotating case remains to
be found. It likewise would be preferable to obtain a clear understanding
of the conserved mass for the $\Lambda \geq 0$ cases, since standard
holographic methods do not directly apply. It would also be interesting to
see how changes in the boundary metric affect the shape of these AdS
black holes \cite{Horowitz:2018coe}.

\subsection*{Acknowledgments}
\label{sc:acknowledgements}
We would like to thank Michael Appels, Pavel Krtou\v{s}, Rob Myers,
and Kostas Skenderis for useful discussions.
This work was supported in part by the Alexander von Humboldt
foundation (AA), a Chilean FONDECYT [Grant No.\ 1181047 (AA),
1170279 (AA), 1161418 (AA)],  CONYCT-RCUK [Newton-Picarte
Grants DPI20140053 (AA) and DPI20140115 (AA)], by the STFC
[consolidated grant ST/P000371/1 (RG)], the Natural Sciences and Engineering
Research Council of Canada (DK and RBM), and by the Perimeter Institute (FG, DK
and RG).
Research at Perimeter Institute is supported by the Government of Canada
through the Department of Innovation, Science and Economic Development
and by the Province of Ontario through the Ministry of Research and Innovation.
RG would also like to thank the Simons Foundation for support and the
Aspen Center for Physics for hospitality while this work was being completed.

\appendix

\section{Holographic stress tensor}
\label{app:hol}

In this appendix we provide further detail, and gather the formulae for
the holographic stress tensor and boundary metric variations, concentrating
mainly on the slowly accelerating charged (non-rotating) AdS C-metric, as
the expressions once rotation is included are extremely cumbersome.

The first step is to write a coordinate transformation between the C-metric
coordinates in \eqref{eq:metric} and the FG coordinates of \eqref{FGmetric}
in a series expansion\footnote{Note that the $r-$coordinate
in \eqref{eq:metric} only covers half of the boundary, since $r\to\infty$
at $\theta=\pi/2$, therefore we use $y = 1/r$ as an alternate radial
coordinate.} as $z\to0$:
\begin{equation}\label{FGt}
\frac{1}{r}= -Ax-\sum_{n=1}^{4}F_{n}\left( x\right)z^n\,, \quad
\cos \theta = x+\sum_{n=1}^{4}G_{n}\left( x\right)z^n\, .
\end{equation}
One inserts this expansion
\eqref{FGt} into \eqref{eq:metric} and inspects the terms order by order in $z$.
The functions $F_n$ and $G_n$ are then found
by requiring that $g_{zz} \equiv \ell^2/z^2$, and $g_{za} \equiv 0$
at each order in $z$; this determines all but the function $F_1$, which
appears as an overall conformal factor as expected. The
$\gamma^{(n)}_{ab}$ are also determined by these expressions.

To illustrate the process, note that
\be
g_{zx} = 0 \quad \Rightarrow \quad \frac{1}{h} \frac{\partial \theta}{\partial z}
\frac{\partial \theta}{\partial x} = - \frac{r^2}{f}
\frac{\partial r^{-1}}{\partial z} \frac{\partial r^{-1}}{\partial x}\,,
\ee
which implies
\be
\beal
&\frac{1}{\ell^2} \left (\sum n G_n z^{n-1} \right )
\left ( A + \sum G_n' (x) z^n \right) \Upsilon_a^2
\left [x + \sum \frac{F_n}{A} z^n \right ]
\qquad \qquad \\
&\qquad \qquad \qquad = -
\left ( \sum n F_n z^{n-1} \right)
\left ( A + \sum F_n' (x) z^n \right)X\left [x + \sum G_n z^n \right ]\,,
\eeal
\label{expansion}
\ee
where
\be
\beal
X[\xi] &= (1-\xi^{2})\left( 1+2mA\xi+(e^2+a^2)A^2\xi^2-a^2 \xi^2 /\ell^2\right)\,,\\
\Upsilon_a[\xi] &= \sqrt{1+ a^2 A^2 \xi^4-A^2\ell^2 X(\xi)}\,,
\eeal
\ee
have been written to keep notation concise.
A systematic expansion, order by order, then determines the $G_n$
in terms of the $F_n$. For example,
at leading order, ${\cal O}(z^{-2})$, gives
\be
G_1 (x) = - \frac{A \ell^2 F_1(x) X(x) }{\Upsilon^2_a(x)}\,.
\ee
This then is sufficient to ensure $g_{zz} = \ell^2/z^2$ to this order,
and gives the boundary metric as:
\be
\beal
ds_{(0)}^2=
\frac{\Upsilon_a^4}
{\Sigma_{FG}^3 F_1^2}\left( X \left [ a A^2 x^2 \frac{dt}{\alpha} -
(1+a^2 A^2 x^2) \frac{d\varphi}{K}\right]^2
- \frac{\Upsilon_a^2}
{\ell^2}
\left [ \frac{dt}{\alpha} - a (1-x^2) \frac{d\varphi}{K} \right]^2 \right)
+ \frac{\Upsilon_a^2}{F_1^2 X} dx^2\,,
\eeal
\label{gamma0}
\ee
where, again for conciseness, we write
\be
\Sigma_{FG} = (1+ a^2 A^2 x^4)\,,
\ee
and the arguments of all the functions are understood now as being `$x$'.
Note that the transformation \eqref{FGt} is valid in general only when
$\Upsilon^2_a>0$, which precisely translates into the condition that acceleration
horizons are absent, i.e.\ we are working with the slowly accelerating C-metric.

To proceed further, we must now Taylor expand the functions $X$ and
$\Upsilon^2_a$ in \eqref{expansion}, and the expressions rapidly become
rather lengthy for $a\neq0$. Setting $a=0$,
which sets $\Sigma_{FG}=1$, $\Upsilon_a=\Upsilon=  \sqrt{1-A^2\ell^2 X}$, and further
writing $F_{1}(x) = - \Upsilon^3/ \alpha \omega(x)$ in order to
elucidate the conformal degree of freedom in the boundary metric,
$\omega$, allows a more tractable computation of the $\gamma^{(n)}$,
and hence the stress tensor.
In coordinates $(t,x,\varphi)$, the expectation value of the energy
momentum of the CFT$_{3}$ reads
\begin{equation}
\left\langle {\cal T} ^a_b \right\rangle
=\lim_{z\to0 }\frac{1}{\ell z}{\cal T}^a_b
={-}\frac{X'''(x)\Upsilon^3}{{96\pi} A \alpha^{{3}} \omega^3}\, \text{diag}\; \Bigl[
-(2-3A^2\ell^2 X)\,, \; 1\,, \; (1- 3 A^2 \ell^2 X) \Bigr]\,.
\ee
This can be interpreted as an energy momentum tensor of a thermal
perfect fluid with no dissipation terms plus a non-hydrodynamic
correction \cite{deFreitas:2014lia, Anabalon:2018ydc}.
With respect to the static observer at infinity, $U=\omega^{-1}\partial_\tau$,
this yields the following energy density:
\begin{equation}
\rho _{E}= \frac{(m{+2 e^2 A x})}{8\pi\ell^2 \alpha^3\omega^3}\Upsilon^3
(2-3A^{2}\ell^{2}X)\,,
\end{equation}
which upon integration, yields the holographic mass
\begin{equation}
M= \int \rho _{E} \ell^3 \sqrt{-\gamma^{(0)}}~dxd\varphi
=\frac{m(1-A^2\ell^2-A^4e^2\ell^2)}{K\alpha}\,,
\end{equation}
in agreement with \eqref{AdSthermo}.
Finally, we find the variation of the boundary metric with respect to
the parameters,
\be
\delta \gamma^{ab}=\frac{\partial \gamma^{ab}}{\partial K} \delta K
+ \frac{\partial \gamma^{ab}}{\partial A} \delta A
+\frac{\partial \gamma^{ab}}{\partial m} \delta m
+\frac{\partial \gamma^{ab}}{\partial e} \delta e\,,
\ee
keeping $\ell$ and $\mu$'s constant, and calculate
\be
\delta I=\int_{\partial M} \sqrt{-\gamma}\tau_{ab} \delta \gamma^{ab} d^3x\,.
\ee
Imposing that the variation vanishes we find that the unknown
parameter $\alpha$ must be
\be
\alpha=\sqrt{(1+A^2e^2)(1-A^2\ell^2-A^4\ell^2e^2)}\,,
\ee
which agrees with \eqref{alp-eq} in the case $a=0$.

\providecommand{\href}[2]{#2}
\begingroup\raggedright%

\end{document}